\documentstyle[aps,epsfig,preprint]{revtex}

\newcommand{\r}{{\bf r}}
\newcommand{\p}{{\bf p}}
\newcommand{\w}{{\bf w}}
\newcommand\be{\begin{equation}}
\newcommand\ee{\end{equation}}

\begin{document}

\title{Folding Lennard-Jones proteins by a contact potential}

\author{Cecilia Clementi$^{1,\dag}$ \footnote{corresponding author:
e-mail address: cecilia@curio.ucsd.edu, telephone number: +1--619--534--7337,
fax number +1--619--534--7697}, Michele Vendruscolo$^{2,\ddag}$,
Amos Maritan$^1$ and Eytan Domany$^2$}

\address{$^1$ International School for Advanced Studies (SISSA)}
\address{and Istituto Nazionale di Fisica della Materia,}
\address{Via Beirut 2-4, 34014 Trieste, Italy}
\address{and  Abdus Salam International Center for Theoretical Physics,}
\address{Strada Costiera 11, 34014 Trieste, Italy}
\address{$^2$ Department of Physics of Complex Systems, Weizmann
Institute of Science, \\ 
Rehovot 76100, Israel}
\address{$^{\dag}$ Present address:\\ Department of Physics, University of
California at San Diego,\\
La Jolla, California 92093-0319, USA}
\address{$^{\ddag}$ Present address: \\
Oxford Centre for Molecular Sciences, New Chemistry Laboratory \\
South Parks Road, OX1 3QT Oxford, UK }
\address{}
\date{\today }
\maketitle

\noindent{\bf Key words : protein folding problem, off--lattice
model, contact map, \\ Molecular Dynamic simulation, Monte Carlo
simulation.}

\begin{abstract}
{\noindent
We studied the possibility to approximate a Lennard Jones interaction 
by a pairwise contact potential. 
First we used a Lennard-Jones potential to design off-lattice, 
protein-like heteropolymer sequences, whose lowest energy 
(native) conformations were then identified by Molecular Dynamics.
Then we turned to investigate whether one can find a
pairwise {\it contact} potential, whose ground states are the contact maps 
associated with these native conformations. 
We show that such a requirement cannot be satisfied exactly - i.e. no
such contact parameters exist. Nevertheless,
we found that one can find contact energy parameters for which an energy
minimization procedure, acting in the space of contact maps, yields 
maps whose corresponding structures are  close to the native ones.
Finally we show that when these structures are used as the initial point of
a Molecular Dynamics energy minimization process, the correct native
folds are recovered with high probability. }
\end{abstract}

\centerline{ PACS numbers: 87.15.By, 87.10.+e }

\section{Introduction}

One of the most challenging open problems in computational molecular 
biology is the prediction of a protein's structure from its amino acid sequence -- the so called protein folding problem.
Assuming that the folded state is thermodynamically stable \cite{anfinsen}
the problem can be formulated as follows: for given sequence of amino acids
and interactions between them, find the conformation of minimal energy.
A successful prediction is marked by the experimental validation
of the structure.
Such kind of prediction is still unfeasible \cite{f97}.

In order to tackle the protein folding problem from a theoretical point
of view, in practice one has to choose 
1) a representation of protein structure;
2) an approximation for the energy;
3) an algorithm to minimize the energy (i.e. to find the lowest energy
structure for a given amino acid sequence).
A conceptually straightforward approach is to perform
detailed Molecular Dynamics simulations of fully detailed all atom models.
However, numerical integration of Newton's equations is unrealistic 
with present computers on the time-scale of the folding process.
Even more problematic is the choice of the energy function.
An imperfect parameterization of the classical effective 
interactions between atoms could correspond to an energy 
landscape with minima unrelated to the desired ones.

The true energy function which dictates the folding of proteins
is unknown and one must resort to simple approximations.
A fairly commonly used approximation, that of contact energies, was 
recently proved \cite{vd98a} to be too crude to successfully fold {\it real} proteins,
whose native fold is stabilized by the {\it ''true"} potential function.
It is not known whether this failure is
caused by the extreme complexity of the true potential, or to the
oversimplification inherent in the contact energy approximation.
Resolving  this issue is the main purpose of this work. 
To do this, we pose and study in detail a very simple question:
\begin{quote}
{{\em 
Can artificial ``proteins'', 
whose constituent residues interact by a} Lennard-Jones {\em pairwise potential, 
be folded successfully by a pairwise} contact {\em potential ?}
}
\end{quote}
The aim of such a study is to uncover some of the general problems which 
arise when one is forced to resort to energy approximations.

The Weizmann group has developed the contact map approach to 
protein folding.
The contact map \cite{md96,vkd97,vd98a,dill,godzik,holm}
of a protein with $N$ residues is an $N\times N$ matrix ${\bf S}$, 
whose elements are defined as $S_{ij}$=1 
if residues $i$ and $j$
are in contact, and 0 otherwise.
Contact between two residues can be defined in different ways; one is to
consider two amino acids in contact when their two $C_{\alpha}$ atoms are
closer than some threshold (they used 8.5 \AA \cite{vkd97}). Another
definition is based on the minimal distance between two ions that belong to
the two residues \cite{md96,hl94}.
The central premise of this approach is that the contact map
representation has an important computational advantage.
Changing a few contacts in a map induces rather significant large-scale
coherent moves of the corresponding polypeptide chain \cite{vd98b}.

In order to carry out the program they had to solve two problems:
\begin{enumerate}
\item
Finding
an efficient procedure to execute a search in contact
map space \cite{vd98b}
\item
Developing a test to determine whether the resulting maps 
correspond to physically
realizable conformations \cite{vkd97}
\end{enumerate}
Using these techniques,
they have demonstrated that a the simple contact energy approximation
is unsuitable to assign the lowest energy to the native state
even for {\em one} protein. \cite{vd98a}
This last result highlights the fact that the bottleneck in the successful
prediction of protein structure is the choice of the
approximation for the energy.

The Sissa group has pursued a different strategy developing minimalist
off-lattice models of proteins.
By studying toy protein models they were able to reproduce 
the essential features of the real folding process in terms of stability 
and accessibility (first of all the typical 
``funnel'' structure for the energy landscape, that has been shown \cite{wot95} 
to play a critical role).
They found that the case of a $C_{\alpha}$ chain with
4 species of amino acids, equipped by a suitable design procedure,
is effective to represent global features of 
the energy landscape \cite{cecilia}.
In their model, the potential is the simplest generic
off-lattice potential that maintains chain connectivity, provides
an attractive interaction between monomers and ensures self-avoidance:
a polynomial potential between beads that are nearest
neighbors along the chain (to represent the
$C_\alpha$-$C_\alpha$ virtual bond) and a Lennard-Jones (LJ) potential
among all other beads.
This off-lattice toy model of proteins has been a suitable tool to
test potential extraction techniques \cite{smb98,jort} 
and, on the other hand, it useful to study of the dynamical properties can provide 
important insights into the real mechanisms acting in real proteins.

In this work, we investigate whether the contact map representation
is as a suitable approximation to the LJ model of proteins;
that is, whether there exists {\it  a 
suitable choice of contact energy parameters}, for which
the ground states of a pairwise contact potential constitute a good
approximation to the lowest energy states of the
LJ potential. 
We also discuss whether dynamics in contact map space, used in combination
with Molecular Dynamics can, 
at least in the simple case presented here,
be successfully used to perform energy minimization.

\section{Protein Design with Lennard-Jones Potentials}
\label{sec:md_design}
\subsection{Definition of the Off-Lattice Model}

In this first part of the study we represent a protein by a chain of $N$ 
monomers or beads, representing the $C_{\alpha}$ atoms of the amino acids.
A configuration $\r$ of the chain is defined by the coordinates 
$\r_1,\ldots,\r_N$ of each $C_{\alpha}$ atom in  
three-dimensional continuous space. We 
consider effective pairwise energies between amino acids.
The interaction potential $V_{i,j}$ between the monomers $i$ and $j$
is the sum of a covalent bond term plus a LJ term:
\be 
\label{eq:LJ} 
V_{i,j} = \delta_{i,j-1}f(r_{ij}) + (1-\delta_{i,j-1}) V^{LJ}_{ij}
\end{equation}
where 
\begin{equation}
V^{LJ}_{ij}=
\eta_{ij}\left[ 
(\frac{\sigma_{ij}}{r_{ij}})^{12} - 
(\frac{\sigma_{ij}}{r_{ij}})^6\right] , 
\label{LJ_eq.2}
\ee 
and $r_{ij}=|\r_i-\r_j|$ are the interparticle distances. 

For the energy bond function $f(r_{ij})$ we choose the expression: 
\be 
f(x)=a (x-d_0)^2+b (x-d_0)^4, 
\ee 
that is the fourth-order expansion of a generic and symmetric 
function 
of the distance $x$. 
The parameters $a$ and $b$ are taken to be $1$ and $100$ respectively.
We add a quartic term to the usual quadratic one \cite{thiru}
because a plain harmonic potential could
induce energy localization in some specific
modes, significantly increasing the time needed for equilibration.

The parameter $d_0$ represents the equilibrium distance of the 
nearest neighbors along the chain. 
We set $d_0 = 3.8$, as this is the experimental value for the mean
distance between nearest neighbor $C_{\alpha}$ atoms along the 
chain in 
real proteins, as taken from the Protein Data Bank. 

The total energy of a chain in conformation $\r$ is defined as: 
\be 
E=\sum_{i=1}^{N} \frac{\p_{i}^2}{2} 
+ \sum_{i=1}^{N} \sum_{j>i} V_{ij}. 
\ee 
The first term is the classical kinetic energy of the chain, where the 
$\p_{i}$'s are the canonical momenta conjugate to the 
$\r_{i}$'s.

\subsection{First step: generation of off-lattice native-like structures}   
\label{step1} 

Off-lattice there are an infinite number of conformations almost all 
of which are not designable (i.e. there is no sequence which has the 
conformation as its ground state).
We aim at working with dynamically accessible, designable and 
compact structures. We present here an outline of the procedure;
further details can be found in Ref. \cite{cecilia}.
The simplest way to obtain such configurations is to initially  deal 
with homopolymer chains, setting the parameters $\eta_{ij}=\eta$ and 
$\sigma_{ij} = \sigma$ of Eq.~(\ref{LJ_eq.2}) for all the beads in the chain. 
By cooling such a chain 
below the $\theta$ temperature \footnote{It is known \cite{Flory}
that varying the temperature an homopolymer chain presents two different
phases according to the dominance of attractive or repulsive interaction
energy (as it can be with the LJ potential). The transition temperature is usually
called $\theta$ temperature.} we obtain maximally compact configurations. 

We fix the parameters to the constant values $\eta = 40$ and $\sigma = 
6.5$. The parameter $\eta$ determines the energy scale, while $\sigma$ 
determines the interaction length between monomers. Such a value for 
$\sigma$ ensures that, in practice, 
two monomers significantly attract each other 
for distances smaller then $8$ \AA, as we show in \S \ref{sqarewell}.
Such distance threshold is 
usually used for the inter-amino acids bond, and is determined by the 
requirement that the average number of $C_{\alpha} - C_{\alpha}$ 
contacts for each amino acid is roughly equal to the respective numbers 
obtained with the all-atom definition of contacts 
\cite{md96}. 

We used Molecular Dynamics (MD)
(integrating Newton's
laws of motion on a computer) for simulating the kinetics of the
chains.
We employed an efficient and precise simplectic algorithm
\cite{LapoAlgo}, in which one varies the
energy density $\epsilon=E/N$, which is related to the temperature
\cite{PLV}.
In order to find low energy structures of the  chain, we used a 
combination of MD  and slow cooling. 
Starting 
with different, randomly selected, initial conditions we find different 
low energy configurations for the chain, since the minimum energy state of 
a homopolymer is 
largely degenerate. 
For each MD simulation the chain is equilibrated 
at its initial energy during 8000 integration steps. We estimate 
the temperature of the system by computing the average value of the 
kinetic energy of the chain \cite{PLV}. 
Then, after 8000 steps, we slowly decrease the temperature by rescaling
each component of the momenta by the same factor ( $< 1$). 
We use the value $0.8$ as a cooling factor. 
Then the chain is again equilibrated at the new temperature and the 
procedure is iterated until very low temperatures are reached.
At low temperature the chain is ``trapped'' in the compact configuration 
of minimum energy. 

We  performed $6$ times this algorithm and  obtained 
$6$ different native-like structures for sequences of lengths 
$N=30$, whose coordinates
will be denoted by $\r_\alpha^*$, ($\alpha=1,\ldots,6$).

We measure the difference between two structures of equal length N, 
using the root mean square (RMS) distance $D$: 
\be 
D(\r_i,\r_i^\prime) = \sqrt{\frac{1}{N} \sum_{i=1}^{N} (\r_{i}-\r_i^\prime)^2} 
\label{dist} 
\ee 
where one structure is translated and rotated to get a minimal 
$D$. The standard procedure, described in  Ref. \cite{kab} was used. 
Two structures are considered different if their distance $D$ is 
larger than $1$ \AA. 
The mean distance among our 6 different native structures is about $5$ \AA. 
\subsection{Second step: protein design}                 
\label{step2} 

At this point we need to select, for each chosen compact structure, 
a sequence having its energy minimum on this structure. 
Furthermore, it should be able to reach the selected structure in 
an accessible time. 
In order to generate such sequences we use the design 
procedure of Ref. \cite{sg93}. 
We use $4$ types of amino-acids,
thus we deal with $10$ different parameters $\eta_{ij}$ ($i,j=1,\ldots,4$). 
We set these parameters by hand, with the
constraint that the inequalities
$2 \eta_{\alpha\beta} < eta_{\alpha\alpha} + eta_{\beta\beta}, \;\;\;
(\alpha\neq \beta)$, are satisfied. The following values have been used: \\
$\eta_{1,1} = 40$, $\eta_{1,2}=\eta_{2,1}= 30$,
$\eta_{1,3} = \eta_{3,1}=20$, $\eta_{1,4}=\eta_{4,1}= 17$,
$\eta_{2,2} = 25$, $\eta_{2,3}=\eta_{3,2}= 13$,
$\eta_{2,4} = \eta_{4,2}=10$, $\eta_{3,3}= 5$,
$\eta_{3,4} = \eta_{4,3}=2$, $\eta_{4,4}= 1$.

We fix an identical amino acid composition for all the sequences 
and the values for each matrix element $\eta_{ij}$. 
Then we select the 
optimal sequence for a target structure using a Monte Carlo 
algorithm in sequence space. Starting with a random sequence, at
fixed composition, we perform random permutations accepting or 
rejecting new sequences with respect to the Boltzmann factor 
$P=\exp(-\Delta E/k_B T)$, where $\Delta E$ is the energy 
variation due to permutation and $T$ is the ``temperature'' 
parameter of the Monte Carlo optimization scheme ($k_B$ is set equal to 1). 
By slowly lowering
the parameter $T$ we select the sequence that has the lowest 
energy in the target conformation. In such a way we modify the 
energy landscape: out of the multidegenerate minimum energy scenario 
of the homopolymer we select a single minimum configuration lowering its
energy
by designing an appropriate sequence. 

This method to select the 
sequence cannot guarantee that the target 
structures are indeed global energy minima. 
To check the obtained sequences we slowly cool (as in \S \ref{step1}) 
each selected sequence
about 50 times (each time starting from a different initial condition). 
The cooling simulations find, as the lowest energy configurations,
the same structures as before.
Hence, these target structures are probably
the lowest energy structures for the selected sequences, at least 
among the dynamically accessible ones, and this result is confirmed
by the simulation in contact map space presented below.
This repeated cooling process generates also a set of alternative, 
higher energy, metastable configurations (typically $2-3$ for each 
sequence) that, if perturbed, ``decay'' to the corresponding 
global minimum configuration. 

\section{Derivation of a set of pairwise contact energy parameters}
\label{sec:cm_enset}

In the {\it pairwise contact approximation} the energy is written as
\begin{equation}
E^{pair} ({\bf a}, {\bf S}, {\bf w}) =
\sum_{i<j}^N {{\bf S}_{ij}}  {w(a_i,a_j)} \; .
\label{eq:pair}
\end{equation}
We denote by ${\bf a}$ the sequence of amino acids,
by ${\bf S}$ the conformation (represented by its contact map \cite{vd98a})
and by ${\bf w}$ the set of energy parameters.
If there is a contact between residues $i$ and $j$, 
the parameter $w(a_i,a_j)$,
which represents the energy gained by bringing amino acids
$a_i$ and $a_j$ in contact, is added to the energy.
Two amino acids are said to be in contact if their $C_\alpha$ atoms are closer
than an upper threshold distance $R_U$ whose value will be discussed
below.

We now explain in detail the approximation involved in Eq. (\ref{eq:pair}).
Denote by ${\cal C}$ a micro-state of the system.
In general the micro-state is specified by the coordinates of all the
atoms of the proteins and of the water molecules of the solvent.
In the present case, we are considering a Lennard-Jones model of a $C_\alpha$
chain, so a micro-state is completely specified
by the coordinates of the N monomers comprising the chain.
Since many microscopic conformations share the
same contact map {\bf S}, it is appropriate to
define a {\it free energy} ${\cal H} ({\bf a},{\bf S})$
associated with this sequence and map: 
\begin{equation}
{\rm Prob}( {\bf S} ) \propto e^{- {\cal H}({\bf a},{\bf S})} 
= \sum_{{\cal C}} e^{-E({\cal C})/k_BT} 
\Delta({\cal C },{\bf S})  
\label{eq:free}
\end{equation}
where 
$\Delta({\cal C},{\bf S})$=1 
if ${\bf S}$ is consistent with ${\cal C}$ and $\Delta=0$ otherwise; 
i.e.  $\Delta$  is a ``projection operator'' which
ensures that only those configurations whose
contact map is {\bf S} contribute to the sum (\ref{eq:free}).
For real proteins $E(\cal{C})$ is the unknown ``true'' microscopic  energy,
here it is the Lennard-Jones energy as of Eq. (\ref{eq:LJ}).

Since it is impossible to evaluate
this  {\it exact} definition of the free energy of a map, we resort
to a phenomenological approach,
guessing the form of 
${\cal H} ({\bf a},{\bf S})$
that would have been obtained had the sum (\ref{eq:free})
been carried out. $E^{pair}({\bf a},{\bf S},\w)$ of eq.
(\ref{eq:pair}) is the simplest approximation to the true free energy. 
To test the extent to which this approximate form
is capable of stabilizing the native map of a protein against other
non-native maps, we must specify the parameters $\w$.

To derive the set ${\bf w}$ of pairwise contact 
energy parameters we require that the following
set of conditions  be satisfied for all the proteins in the database
\cite{vd98a,smb98,jort}
\begin{equation}
E^{pair}({\bf a}, {\bf S}^*, {\bf w}) < 
E^{pair}({\bf a}, {\bf S}_\mu, {\bf w}) \;.
\label{eq:optimization}
\end{equation}
Here ${\bf S}^*$ is the contact map of the native conformation
of sequence ${\bf a}$ and ${\bf S}_{\mu}$ (${\mu}=1,\ldots N_D$)
are contact maps taken from a set of $N_D$ alternative conformations.
The problem is to find a set ${\bf w}$ such that for each decoy $\mu$
the inequality (\ref{eq:optimization}) is satisfied.
The perceptron learning technique \cite{rosenblatt,minpap,km87} is used
to search for contact energy parameters ${\bf w}$ for which
the set of conditions (\ref{eq:optimization}) are satisfied.
The set ${\bf w}$ that gives the optimal solution of this problem
is equivalent to the maximization of
the gap between the ground state and the first excited state, as it is
shown in the Appendix.

\subsection{Generation of alternative conformations}

To obtain a large ensemble of decoys we performed
dynamics in contact map space collecting maps along the trajectory.
The method is presented elsewhere \cite{vd98b};
here we outline only the details that are relevant for the present
implementation.
Our algorithm is divided in four steps.
\begin{enumerate}
\item
We start from an existing map and perform large scale
``cluster'' moves.
At this stage, no attempt is made to preserve physicality.
The contact map which is obtained by this procedure is typically
uncorrelated to the starting one.
\item
The resulting map is refined by using local moves.
\item
We use the reconstruction algorithm \cite{vkd97}
to restore physicality by projecting the map obtained from the second step
onto the physical subspace.
\item
We perform a further optimization
by an energy minimization
in real space using a standard Metropolis crankshaft technique
\cite{vd98b}.
\end{enumerate}

Using this algorithm,
after the choice of a suitable definition of contact --as
detailed in \S~\ref{subsec:physicality} and \S~\ref{subsec:learnability}--,
we generated a set of $N_D$=60000 alternative conformations, 10000 for each 
of the 6 sequences in the database.
As discussed in Ref. \cite{vd98b},
the contact maps that are obtained by contact map dynamics are uncorrelated
and the $N_D$ decoys form a representative set of low energy
competitors for the native state.
The important requirement is to generate a large set of uncorrelated
decoys, since a good energy function must stabilize the correct structure
and destabilize all the others.

This selection of candidates was performed using
the set of LJ parameters $\eta_{ij}$ as a first guess
for the pairwise contact energy parameters ${\bf w}$.

\section{Square well approximation of the LJ potential}
\label{sqarewell}

\subsection{Truncation of the LJ potential}

In the definition of contact one has to specify the upper threshold $R_U$
of the distance between two beads,
below which  are considered to be in contact.
To get an estimate of a reasonable value of $R_U$ we studied the truncation
of the LJ potential.
Using MD, we analyzed the stability of the folded
conformations against a cut in the tail of the LJ potential. We used
a potential equal to the Lennard-Jones function
for $r$  {\em smaller} than
$R_C$, and equal to zero for $r$ larger than $R_C$:
\begin{equation}
\label{eq:ljtrunc}
V_{LJ}^{\prime} (r_{ij})  
 = \left\{
\begin{array}{ll}
 \eta_{ij}(
(\frac{\sigma}{r_{ij}})^{12} -
(\frac{\sigma}{r_{ij}})^6)
 &
 {\rm if}\;\; r_{ij}  < R_C \\
 & \\
~0
 &
 {\rm if}\;\; r_{ij}  > R_C \\
\end{array}
\right.
\end{equation}
For a given $R_C$, we determined the ground state conformations 
of the 6 proteins in the database.
This was done by performing $n$ runs of Molecular Dynamics 
minimization (by slow cooling, as described in \S \ref{step1}
with typically $n=50$).
For each protein, each minimization ended in a slightly different
structure $\r_\alpha^\nu(R_C)$ ($\alpha=1,\ldots,6$, $\nu=1,n$)
We measured the RMS distance $\langle D(R_C) \rangle$, 
averaged over the 6 proteins in the database,
between the native conformations and the conformations found as 
minima of the potential of Eq. (\ref{eq:ljtrunc}):
\begin{equation}
\langle D(R_C) \rangle = \sum_{\alpha=1}^6 \sum_{\nu=1}^n 
D(\r_\alpha^*,\r_\alpha^\nu(R_C)) \;.
\end{equation}
In Fig. \ref{fig:cut} we show the results for each protein, for several choices of
the cutting length $R_C$. In the inset the average distance $\langle D(R_C) \rangle$ 
with its standard deviation as a function of the cutting length $R_C$ are shown.
Remarkably, it is possible to cut the LJ potential inside the
attractive well down to $R_C$=8 \AA, still keeping the average distance
below 1 \AA.

\subsection{Derivation of the region of physicality}
\label{subsec:physicality}

For any  value of $R_U$ the contact map ${\bf S}$ of any chain conformation
can be easily determined. In order to construct decoys, however, we must deal with
the inverse problem --
to get a chain conformation starting from some contact map ${\bf S}$. 
In order to apply the reconstruction procedure 
introduced in Ref. \cite{vkd97} for a possibly non physical contact map
one has to specify another length; the hard core distance $R_L$, defined
as the minimal allowed distance between two amino acids. This is essential
since without such a lower
cutoff we would be able to construct chain configurations for maps with
much higher numbers of contacts 
than the native maps and hence would identify such dense decoys as physical. 
Since the contact energies tend to be negative such decoys will have much lower 
energies than the native map and will never be able to find a solution to the
inequalities eq. (\ref{eq:newopt}).

In Fig. \ref{fig:dist.d} we show the histogram of all the pairwise distances
for the six native conformations considered in this study.
Apart from the peak at 3.7 \AA, due to consecutive residues along the chain,
the distances range from 6.1 to 18.8 \AA.
The vertical line indicates 6.9 \AA, a distance whose role will be discussed below.
There are 2436 distances between non-consecutive beads,
of which 49 are below 6.9 \AA \hspace{1pt} and 7 below 6.5 \AA.

We now consider the plane $(R_U,R_L)$ and set about to determine
the ``region of physicality'' which is that part of the plane whose
points allow reconstruction of the contact maps that correspond to the 
six native folds. To understand the problem we are facing, one should bear in 
mind that once we chose a value for $R_U$, the contact map associated
with a conformation is determined. When, however, we try to reconstruct one of
these maps, we must also specify a value for $R_L$ and if the chosen value is 
too high, we may find that no corresponding chain configuration exists. In this
case, according to definition, the native contact maps are {\it non-physical}.
The gross features of such a
region can be outlined by some simple arguments. 
First, since we must have $R_U>R_L$, 
the region of physicality must clearly 
lie below the line $R_U=R_L$. 
Second, if $R_L$ is too high,
the contact maps
are non physical and there must be a (possibly horizontal)line that bounds the 
physical region from above.
The exact location of this line 
must be determined numerically.
For $R_L<6.1$ it is clear that all the native contact maps are physical, but
for $R_L>6.1$ \AA \hspace{1pt} the situation is not clear.
It may be able  to ''catch" the few ``close'' contacts (those with $d<6.9$ \AA)
by locally stretching the chain. We  proved numerically that
this is indeed possible up to about $R_L<6.7 $ \AA,
as shown in Fig. \ref{fig:region}a and in more detail in
Fig. \ref{fig:region}b. The physical region is the shaded 
trapezoid open on its right side; for points in this region 
we are able to reconstruct chain conformations for all the six 
native maps
(derived from their ''true" chain conformations by the choice of $R_U$).

\subsection{Derivation of the region of learnability}
\label{subsec:learnability}

We turn now to determine a second region in the $(R_U,R_L)$ plane, 
that of `` learnability''. At
a point $(R_U,R_L)$ in this region
it is possible
to find a set of energy parameters ${\bf w}$ such that each of the six native maps
have the lower  energy than all their respective decoys.
As was done above for the region of physicality, it is possible to predict
the general shape of the region of learnability a well.
First, it is limited by the same bisecting line $R_U=R_L$.
Second, there is a vertical line at $R_U=6.1$ \AA,
at the left of which all the contact maps are empty 
-- we are not interested in such a case.
Third, there is a second vertical line, at $R_U=19$, at the right of which
all the contact maps are filled. Also this case is not interesting.
Fourth, there must be a line (approximately horizontal) below which
the energy parameters are unlearnable.
We expect this since for small values of $R_L$ amino acids are
allowed to be very close to each other. Very compact conformations
are possible and since the interaction parameters are on the average 
attractive, such compact conformations have very low energy.
The exact location of this line must be determined numerically.
The result is shown in Figs. \ref{fig:region}a and \ref{fig:region}b.
For each choice of $R_U$ and $R_L$, we generated by contact map dynamics
a set of $N_D=60000$ decoys, 10000 for each sequence in the database.
Using the perceptron algorithm, we determined if such set of decoys
was learnable or not. The outcome of the procedure is the triangular
shaded region in Fig. \ref{fig:region}a, whose lower edge is shown
in detail in Fig. \ref{fig:region}b.

The most interesting result is that the region of physicality
and the region of learnability {\bf do not overlap}.
It is not possible to choose $(R_U,R_L)$ such that the contact maps
of the native LJ conformations in the database are physical
and they can be stabilized by a pairwise contact potential.

Two observation must be made.
First, the points labeled as ``unlearnable'' are rigorous, whereas
those labeled as ``learnable'' are tentative -- it is possible
that by increasing the number of decoys also those points will
become unlearnable. The lower edge of the region of learnability
could actually be moved upwards. The same effect should be expected
by increasing the number of proteins in the database.
Second, the points labeled as ``physical'' are rigorous, whereas
those labeled as ``non physical'' are tentative -- we could not
reconstruct the corresponding contact maps, but this failure could
be due to our reconstruction algorithm. Thus, the upper edge 
of the region of physicality could be moved upwards.
However, by increasing the number of proteins in the database
we expect that this upper edge would move downwards.
In the limit of a large number of proteins the latter effect could
possibly dominate.
In conclusion, we argue that the gap between the region of learnability
and the region of physicality is expected to widen when many proteins are 
taken into account.

\section{Folding in contact map space}
Since we have proved that the two regions, 
of learnability and physicality of the native LJ contact maps do not overlap, 
we can conclude that the answer to the question
posed in the Introduction is negative -- 
the contact energy approximation is 
unable to stabilize the native folds of LJ chains. The contact map approximation 
is too crude also for extremely simplified potential as the LJ potential used in 
this work. 

The situation is, however, more subtle.  We can look 
for  contact energies that give rise to
{\it approximate solutions} of the folding problem: it is impossible recover 
the LJ native configurations using the contact map approximation, but we want 
investigate how "close" to the solution the contact map approximation can lead.

To tackle this point, we selected a working point in the learnable region.
We chose $R_U$=8 \AA, according the the minimal value of $R_C$ determined
in Sec. IVA and $R_L=6.9$ \AA. 
As shown above, for this value of $R_L$, native maps are non physical.
In other words, for a given native map
there is no chain which can realize that contact map.
We found, however, that for these $R_L$ and $R_U$
it {\it is} possible to reconstruct conformations whose maps are physical and 
typically differing by only a few contacts from the native map.
Moreover, the corresponding 3D structures are, on the average,
at a RMS distance of less that 2 \AA \hspace{1pt} 
from the native LJ conformations.
We decided to choose $R_L=6.9$ \AA, as the best possible compromise between  
solving Eq. (\ref{eq:optimization}) and our aim is that some {\it physical
maps of low energy} are close to the native maps. 

Now we turn to answer the following important question:
is the set ${\bf w}$ obtained this way useful to fold LJ proteins?
The first aspect of this general question concerns the feasibility
of using contact map dynamics.
That is, we ask:
\begin{quote}
{\em 
Is it possible to fold the sequences in the database 
using contact map dynamics?
}
\end{quote}
Fully successful folding in contact map space
corresponds to exact recovery of the native map.
Since for $R_L=6.9$ \AA  \hspace{1pt} and $R_U=8.0$ \AA \hspace{1pt} the
native contact maps are non physical, such an exact solution is 
not within our reach. Nevertheless, the contact map dynamics could select 
configurations slightly different from the native ones.

For each of the 6 sequences we 
started from some random contact map and ran the contact map dynamics procedure
for 1000 steps; this took about 3 cpu hours on our HP812 computer.
We now analyze in detail our results.

As already observed in Ref. \cite{md96}, the correlation between
energy and distance from the native map
(the ``funnel'') is not always very strong when using $E^{pair}$.
A case of quite good correlations, obtained by contact map dynamics
for sequence N$^o$ 5 (among the six used in this work), is shown in Fig. \ref{fig:corr}.
Since the lowest energy contact map found in the simulation
might or might not coincide with the closest in distance to the native one,
in general we can obtain only a short list of candidates for the native
state. 

For sequence N$^o$ 5  
the physical map of lowest energy is, in fact, the closest in distance to the 
native one. Both maps are shown in Fig. \ref{fig:mseq5_pred}.
The energy $E^{pair}$
of the native map is -35.75 and the energy of the predicted one
is -33.33. The RMS distance between two corresponding conformations, 
shown in Fig. \ref{fig:mseq5.bbc}, is 2.1 \AA.
We emphasize again the obvious fact that since our ${\bf w}$ is a solution of
(\ref{eq:newopt}), 
the native map has lower energy than all the low-energy decoys
that were used for learning. Indeed,
in no case have we found by contact map dynamics
maps lower in energy than the native one. 
We summarize in Table \ref{tab:folding} the results obtained for the six
sequences.

The Hamming distance $D_H$ between two contact maps 
${\bf S}$ and ${\bf S^\prime}$ is defined by
\begin{equation}
D_H = \sum_{j>i} \frac{| S_{ij} - S_{ij}^{\prime} |}
                      {N_c({\bf S})N_c({\bf S}^\prime)} \;,
\label{eq:hamming}
\end{equation}
which counts the number of mismatches between maps
${\bf S}$ and ${\bf S}^\prime$;  $N_c({\bf S})$ and
$N_c({\bf S}^\prime)$ are the numbers of contacts in the two maps.


The second and complementary question we pose is
\begin{quote}
{\em Are the low energy conformations 
generated by contact map dynamics 
good starting points for Molecular Dynamics minimization?}
\end{quote}
We analyze again sequence N$^o$ 5. 
We randomly selected 10 conformations among the 100 of lowest $E^{pair}$ energy
that were generated by contact map dynamics
and reconstructed their corresponding conformations. These were then
used as starting points for a Molecular Dynamics minimization
of the $E^{LJ}$ energy.
In Fig. \ref{fig:md_seq5} we present the energy and distance to the native
conformation for the conformations obtained by MD. As can be seen, 
8 trajectories ended up very close to the native state and 2 in conformations
at RMS distances of 6 and 8 \AA \hspace{1pt} respectively.
We note that these last two initial conformations had the largest RMS distance
from the native state.

We conclude that the predictions performed in contact map space
using the approximate energy $E^{pair}$ are, on the average, 
suitable starting points
for a Molecular Dynamics minimization of the LJ energy.
Moreover, by using Molecular Dynamics it is possible to correctly rank
the predictions that were obtained by contact map dynamics.

\section{Conclusions}
\label{sec:conclusion}


There are four different situations for which
the contact pairwise approximation has been used:

1) When the true potential is a contact pairwise one, such as
for lattice models with nearest-neighbor interactions, 
the energy $E^{pair}$ of Eq. (\ref{eq:pair})
is, by definition, the {\em exact} (free) energy.
Models of this type were investigated in two recent studies \cite{jort,leo}.
In both works, a database of proteins was designed using $E^{pair}$,
with a given choice of energy parameters ${\bf w}^{true}$.
The database was then used to recover a set ${\bf w}$, which  
ideally could be identical to ${\bf w}^{true}$.
Since the native states of the proteins
in the database are the lowest energy conformations of 
$E^{pair}({\bf w}^{true})$,  Eq. (\ref{eq:optimization})
can be satisfied for all the possible conformations for all the sequences
in the database.

2) When the true potential is unknown, such as
for real proteins, Eq. (\ref{eq:pair}) is only an approximation
to the true free energy (\ref{eq:free}), 
and there is no guarantee that such an approximation would lead to good results.
The conformations stabilized by the true energy function
are not necessarily stabilized against all decoys 
by a pairwise contact energy function.
For a particular protein, crambin, 
it has been shown \cite{vd98a} that, in fact, such a solution does not exist
and the contact maps of lowest pairwise energy are not close to the 
native maps.

3) The ``true" potential is a contact pairwise one, 
and it is used to design protein sequences on the structure of 
existing proteins.
This case was considered in Ref. \cite{leo}, where it was shown
that it is unfeasible to design sequences, using a pairwise contact
potential, on structures of existing proteins. 
This result implies that the native structures of existing proteins
are ``designed" in a very peculiar way by the possibly very
complex potential used by Nature. A simple pairwise contact potential
can neither be tuned to stabilize the true sequences (see point 2)
nor used to perform protein design on the true structures.

4) The true potential is a Lennard-Jones pairwise one
and it is used to design sequences on structures typically arising from
Lennard-Jones interactions. The resemblance of such structures
to those of real proteins appear to be reasonable \cite{cecilia},
however, as discussed in this paper, the results obtained for approximate
contact potentials are remarkably different.
Here we have shown 
that it is not possible to stabilize the exact native folds.
This result is somewhat surprising -- the approximation seems 
to be quite reasonable, since both the contact potential and
the LJ potential are characterized by a single length scale. 
However, we also showed that in this case
an approximate way to solve the folding problem can be found.
We showed that a dynamics in contact map space is
a suitable tool to perform energy minimization,
when a correct parameterization of the energy is possible. 
Contact map dynamics provides, for a good choice of contact parameters, 
not just a unique structure corresponding to the lowest energy state, 
but a set of candidates for it. 
When this list of predictions is used as starting configurations
of a MD energy minimization, which
uses the original "true" LJ form of the energy, 
the correct ground states are recovered.

It remains to be investigated if a hybrid method, based on ``fusion"
of contact map dynamics and molecular dynamics, in   
a way that takes advantage of the their respective advantages (that 
in some sense are complementary) may be a powerful tool to fold real 
proteins as well.

\section*{Acknowledgments}
C.C expresses her gratitude for the hospitality to the Weizmann Institute 
where part of this work was carried out.
We thank Devarajan Thirumalai for discussions
and Ron Elber for discussing with us a similar approach,
based on Eq. \ref{eq:optimization} (unpublished).
This research was supported by grants from the Minerva Foundation,
the Germany-Israel Science Foundation (GIF) and by a grant from the
Israeli Ministry of Science. 
During the last three months C.C. has been supported by the
NSF (Grant \# 96-03839) and by the La Jolla Interfaces in Science
program (sponsored by the Burroughs Wellcome Fund).

\begin{appendix}
\section*{Perceptron learning}

We describe here the perceptron learning technique that was used.

We first show that for any conformation the condition 
Eq.(\ref{eq:optimization})
can be trivially expressed as 
\begin{equation}
{\bf w} \cdot {\bf x}^{\mu} > 0
\label{eq:newopt}
\end{equation}
To see this just note that for any map ${\bf S}_{\mu}$ the energy  
(\ref{eq:pair})
is a linear function of the 10 contact energies that can appear and it
can be written as 
\begin{equation}
E^{pair}({\bf a}, {\bf S}_{\mu}, {\bf w} ) =
\sum_{c=1}^{10} N_c({\bf a},{\bf S}_{\mu}) w_c
\label{eq:newener}
\end{equation}
Here the index $c=1,...10$  labels the different contacts that can occur
and $N_c({\bf a},{\bf S}_{\mu})$ is the 
total number of contacts of type $c$ that actually
appear for the sequence ${\bf a}$  in the map ${\bf S}_{\mu}$. 
The difference between the energy of this map and
the native ${\bf S}^*$ is, therefore, 
\begin{equation}
\Delta E_{\mu} = 
\sum_{c=1}^{10} x^{\mu}_c w_c = {\bf w} \cdot {\bf x}_{\mu} 
\label{eq:Ediff}
\end{equation}
where we used the notation
\begin{equation}
 x^{\mu}_c = N_c({\bf a}, {\bf S}_{\mu})- N_c({\bf a},{\bf S}^*)
\label{eq:Ndiff}
\end{equation}
and ${\bf S}^*$ is the native map. 

Each candidate map ${\bf S}_{\mu}$ is represented 
by a vector ${\bf x}^{\mu}$ and
hence the question raised above regarding 
stabilization of ${\bf S}^*$ for a sequence ${\bf a}$ becomes
\begin{quote}
Can one find a vector
$\bf w$ such that condition (\ref{eq:newopt}) holds for all ${\bf x}^\mu$?
\end{quote}
If such
a $\bf w$ exists, it can be found by {\it perceptron learning}.

A perceptron is the simplest neural network \cite{rosenblatt}.
It is aimed to solve the following task. Given  a set $P$ of patterns 
(also called input vectors, examples) ${\bf x}^{\mu}$, $\mu=1,\ldots,P$,
find a vector ${\bf w}$ of weights, 
such that the condition
\begin{equation}
h_{\mu} = {\bf w} \cdot {\bf x}^{\mu} > 0
\label{eq:perceptron}
\end{equation}
is satisfied for every example from the set.
If such a $\bf w$ exists for the training set, the problem is {\it learnable};
if not, it is unlearnable.
We assume that the vector of ``weights'' $\bf w$ is normalized,
\begin{equation}
{\bf w} \cdot {\bf w} = 1
\end{equation}
The vector ${\bf w}$ is ``learned'' in the course of  a training session.
The $P$ patterns are presented cyclically; after presentation of pattern $\mu$
the weights ${\bf w}$ are updated according to the following learning rule:
\begin{equation}
\label{eq:prule}
{\bf w}^{\prime} = \left\{ 
\begin{array}{ll}
\frac{{\bf w} + \eta {\bf x}_{\mu}}
                       {|{\bf w} + \eta {\bf x}_{\mu}|} \qquad & 
{\rm if} \qquad  {\bf w} \cdot {\bf x}_{\mu} <0 \\
 & \\
~ {\bf w} & {\rm otherwise}
\end{array}
\right.
\end{equation}
This procedure is called learning since when the present $\bf w$ misses
the correct ``answer'' $h_{\mu} >0$ for example $\mu$, all weights are 
modified in a manner that reduces the error. No matter what
initial guess for the $\bf w$ one takes,
a convergence theorem guarantees that if a solution ${\bf w}$ exists,
it will be found in a finite number of training steps.
\cite{rosenblatt,minpap}.

If the region of parameter space whose
points are solutions is large, one is interested in the optimal solution.
To obtain the optimal perceptron solution, called the {\it maximal stability
perceptron}, we used the Krauth and Mezard algorithm \cite{km87}.
In this algorithm the condition (\ref{eq:perceptron})
is replaced by
\begin{equation}
h_{\mu} = {\bf w} \cdot {\bf x}^{\mu} > c
\label{eq:msperceptron}
\end{equation}
where $c$ is a positive number that should be made as large as possible.

At each time step the ``worst'' example 
${\bf x}^{\nu}$
is identified, namely the one such that
\begin{equation}
 h_{\nu}={\bf w} \cdot {\bf x}^{\nu} = \min_{\mu} \; {\bf w} \cdot {\bf x}^{\mu}
\end{equation}
and the example $\nu$ is used to update the weights according to the rule
(\ref{eq:prule}).
The field $h_{\nu}(t)$ keeps changing at each time step $t$ and
the procedure is iterated until it levels off to its asymptote.
\end{appendix}

\newpage

\subsection*{ Captions to the figures}

{\bf Fig. 1.}
Upper part: Approximation of the Lennard-Jones potential 
by a contact potential. \\
Lower part: Distance of the conformation
of lowest energy found in a MD simulation, done with a potential 
truncated at $R_C$, to the true native state of the non truncated LJ interaction.
Each dot represents a different run. In the inset the average  
$\langle D(R_C) \rangle$ is shown. Down to $R_C=8$ \AA \hspace{1pt}
the distance to the true native state is below 1 \AA.

{\bf Fig. 2.}
Distance probability distribution between amino acids in the six native
conformations in this study.

{\bf Fig. 3.}
Regions of physicality and of learnability in the ($R_U$,$R_L$) plane.

{\bf Fig. 4.}
Correlation between energy $E^{pair}$
and RMS distance $D$ to the native state
and between energy and the Hamming distance $D_H$ 
(see Eq. (\protect \ref{eq:hamming})). 
Results refer to contact map dynamics.

{\bf Fig. 5.}
Contact maps of sequence N$^o$ 5: 
native (below diagonal) and of lowest energy 
found during the simulation (above diagonal).  

{\bf Fig. 6.}
Native conformation of sequence N$^o$ 5 and its lowest energy
conformation found during the simulation. The RMS distance is 2.1 \AA. 

{\bf Fig. 7.}
MD folding trajectories of sequence N$^o$ 5.
Each trajectory (small connected full dots) is obtained 
starting from 10 conformations (large full dots)
randomly chosen from the 100 conformations 
of lowest energy obtained by contact map dynamics.
Conformations (empty large dots) collected during other MD simulation
(see \S \ref{step2}) are also shown as reference. 

\newpage


\begin{center}
\begin{minipage}{12cm}
\begin{table}
\begin{tabular}{c|c|c|c|c|c|c}
sequence &     1   &      2 &      3 &     4 &      5 &      6 \\ \hline
$E_0$    & -34.78 & -36.39 & -36.56 & -36.29 & -35.75 & -36.17 \\
$E$      & -33.53 & -35.17 & -34.87 & -35.60 & -33.33 & -33.59 \\
$D$      &  4.2   &  4.2   &  2.1   &   2.2  &   2.1  &  4.3   \\
$D_H$    & 0.60   & 0.61   & 0.31   &  0.34  & 0.25   & 0.60   \\
\end{tabular}
\vspace{0.2cm}
\caption{
Summary of the results of the folding experiments in contact map space
on the six sequences, using the pairwise contact parameters of 
maximal stability. $E_0$ is the energy of the native contact map.
$E$ is the energy of the contact map of lowest energy found during 
the simulation, $D$ is the RMS distance to the native state conformation
and $D_H$ is the Hamming distance to the native contact map.
}
\label{tab:folding}
\end{table}
\end{minipage}
\end{center}



\pagestyle{empty}
\thispagestyle{empty}

\begin{figure}
\centerline{\psfig{figure=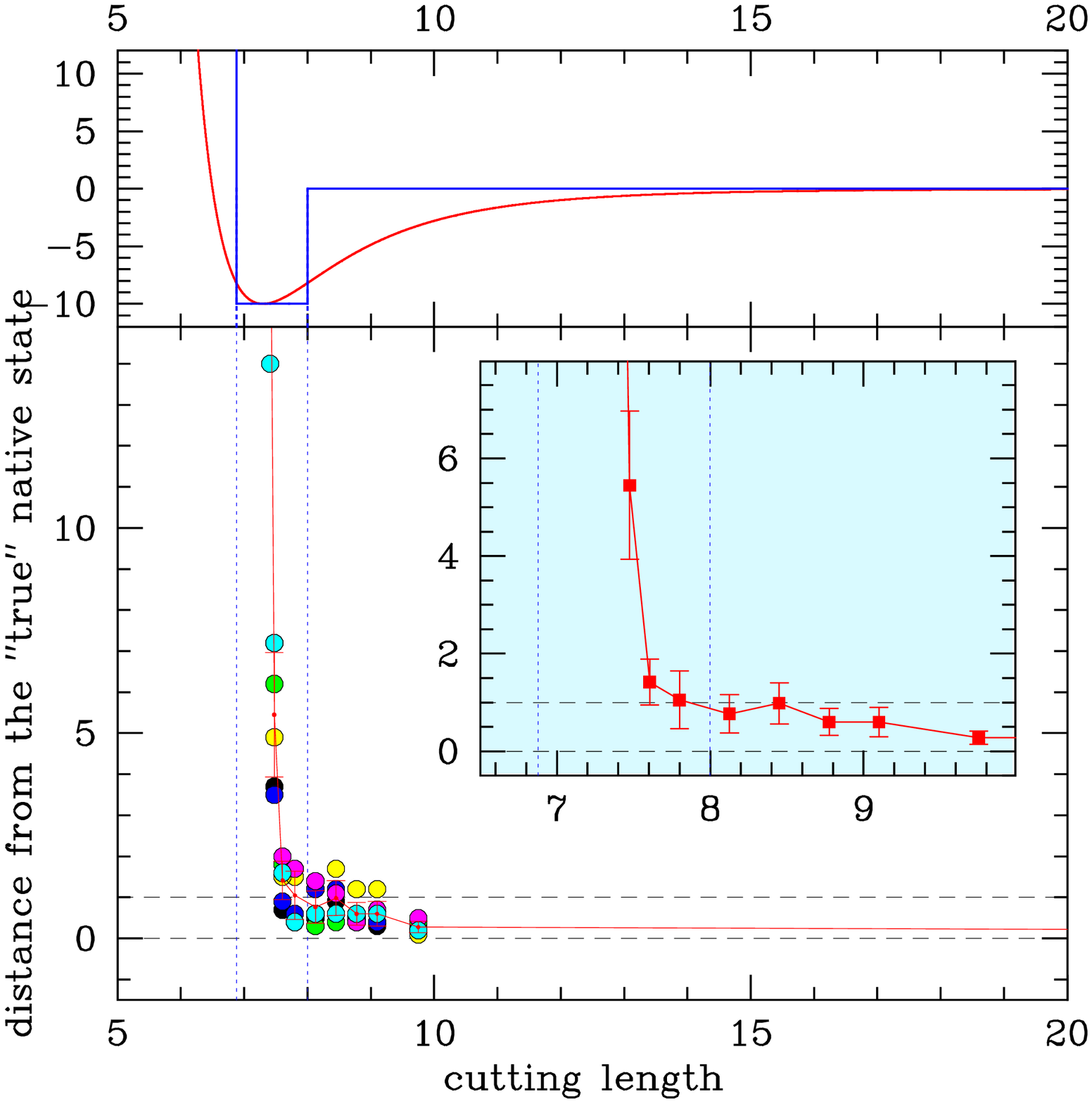,height=7.0cm,angle=0}}
\caption{
}
\label{fig:cut}
\end{figure}

\begin{figure}
\centerline{\psfig{figure=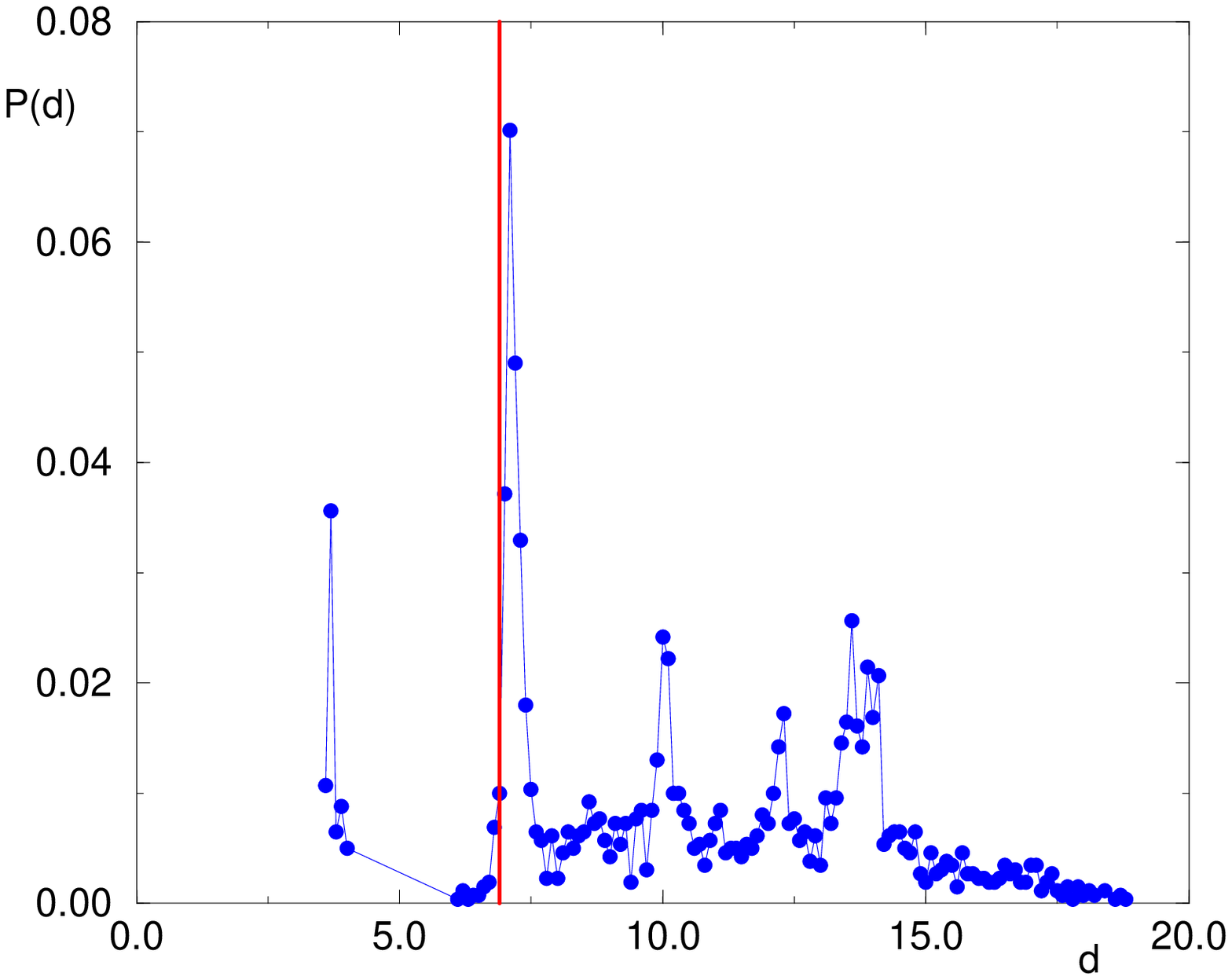,height=7.0cm,angle=0}}
\caption{
}
\label{fig:dist.d}
\end{figure}

\begin{figure}
\centerline{\psfig{figure=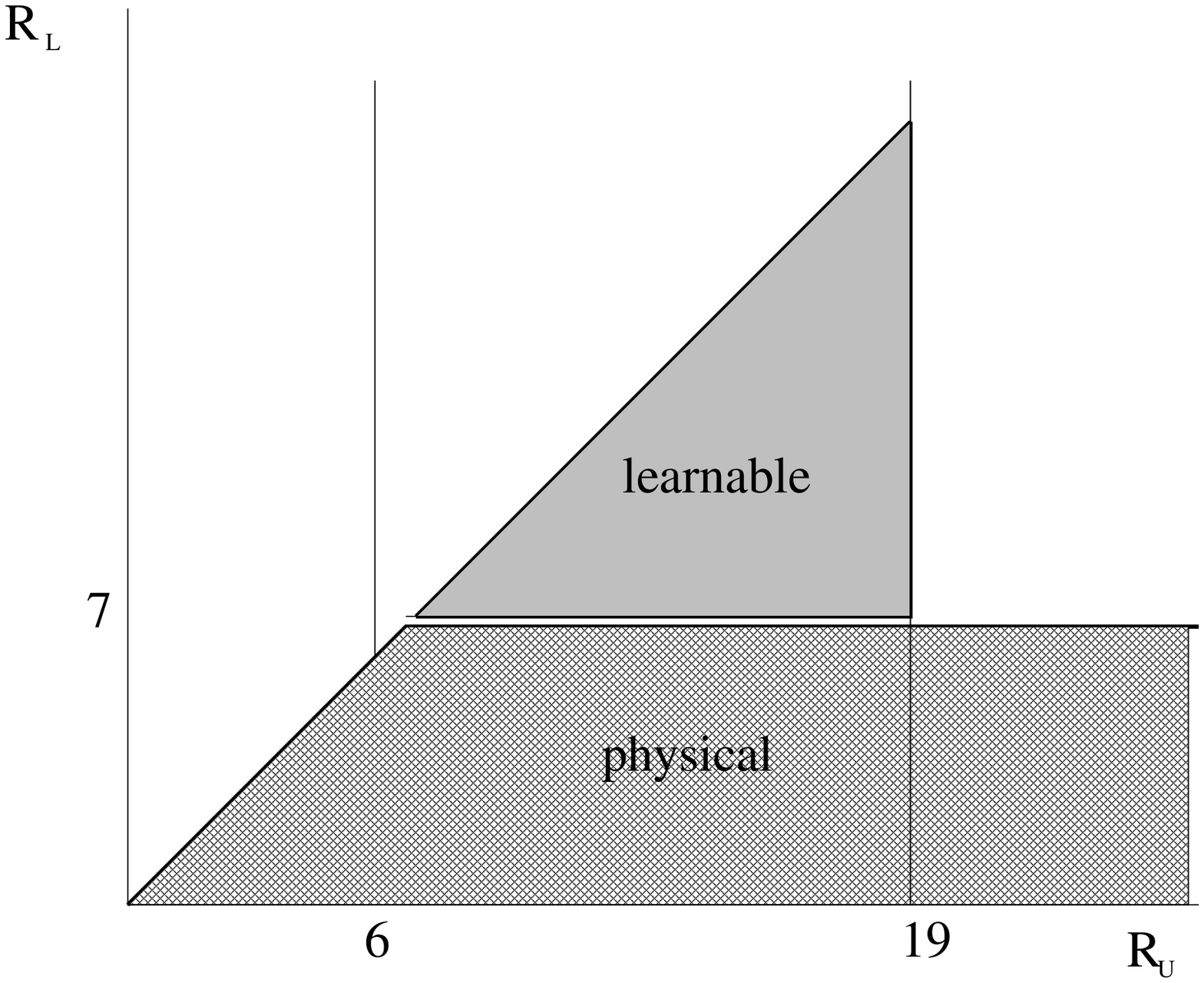,height=6.0cm,angle=0}
            \psfig{figure=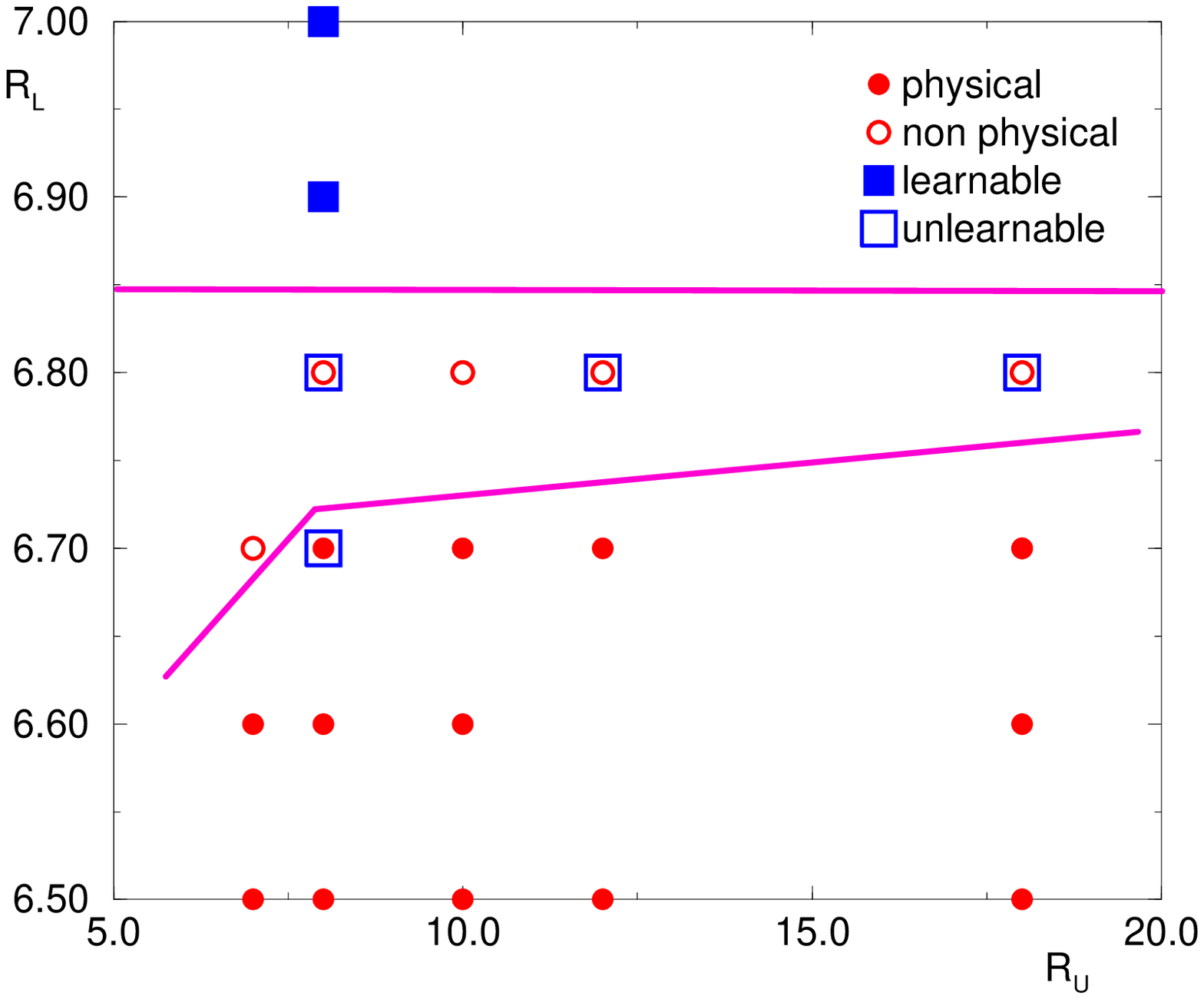,height=7.0cm,angle=0}}
\caption{
}
\label{fig:region}
\end{figure}

\begin{figure}
\centerline{\psfig{figure=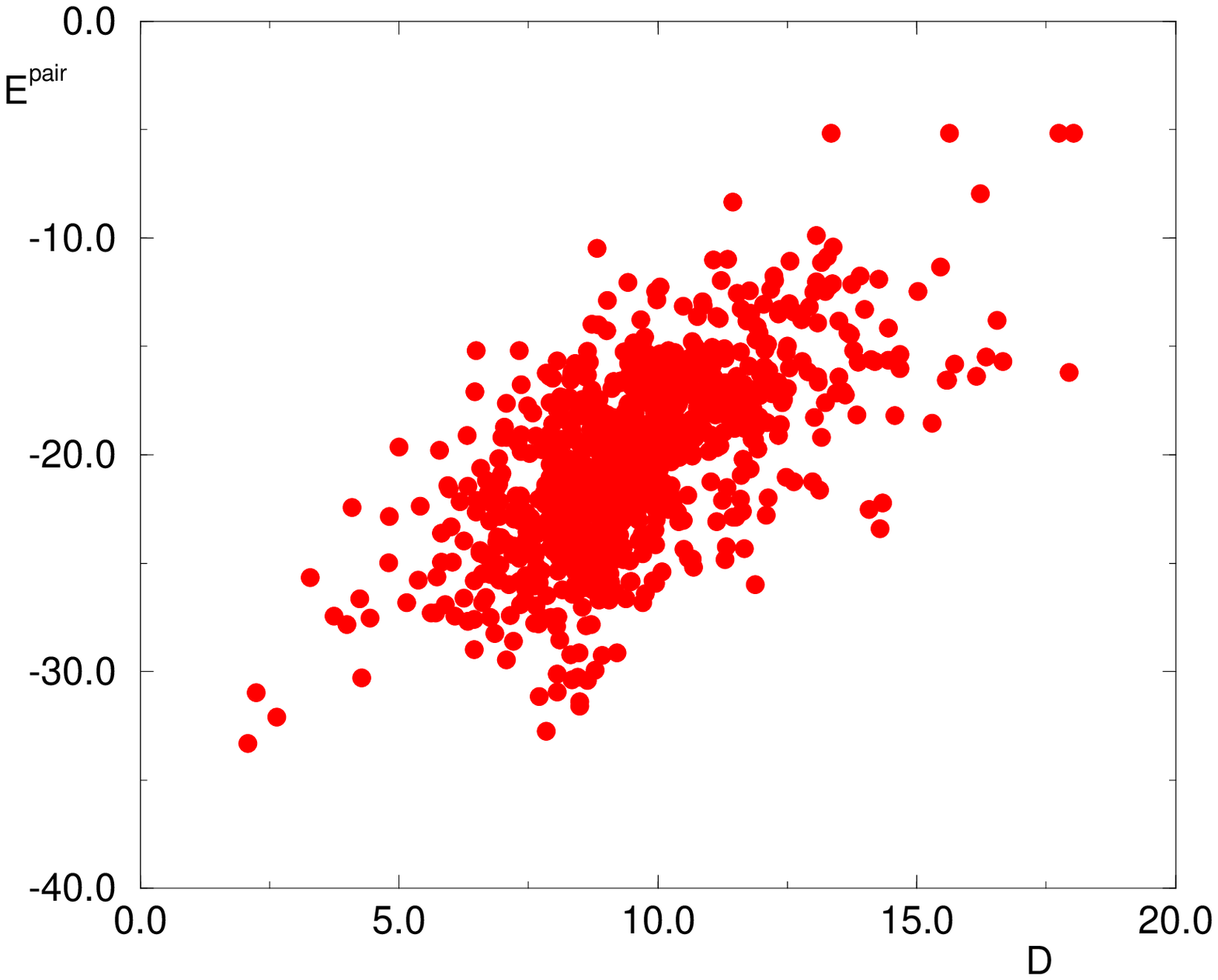,height=7.0cm,angle=0}
            \psfig{figure=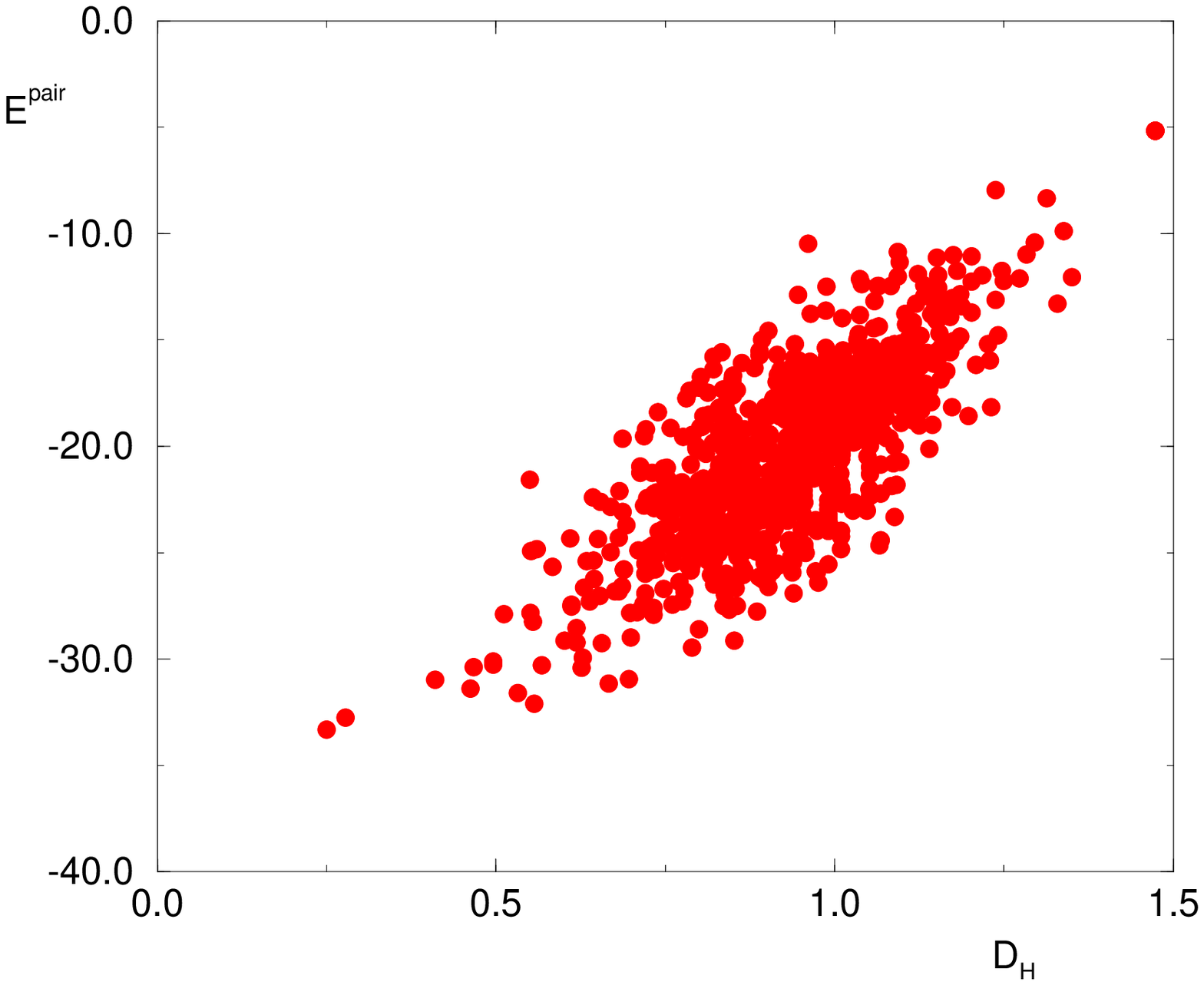,height=7.0cm,angle=0}}
\caption{
}
\label{fig:corr}
\end{figure}

\begin{figure}
\centerline{\psfig{figure=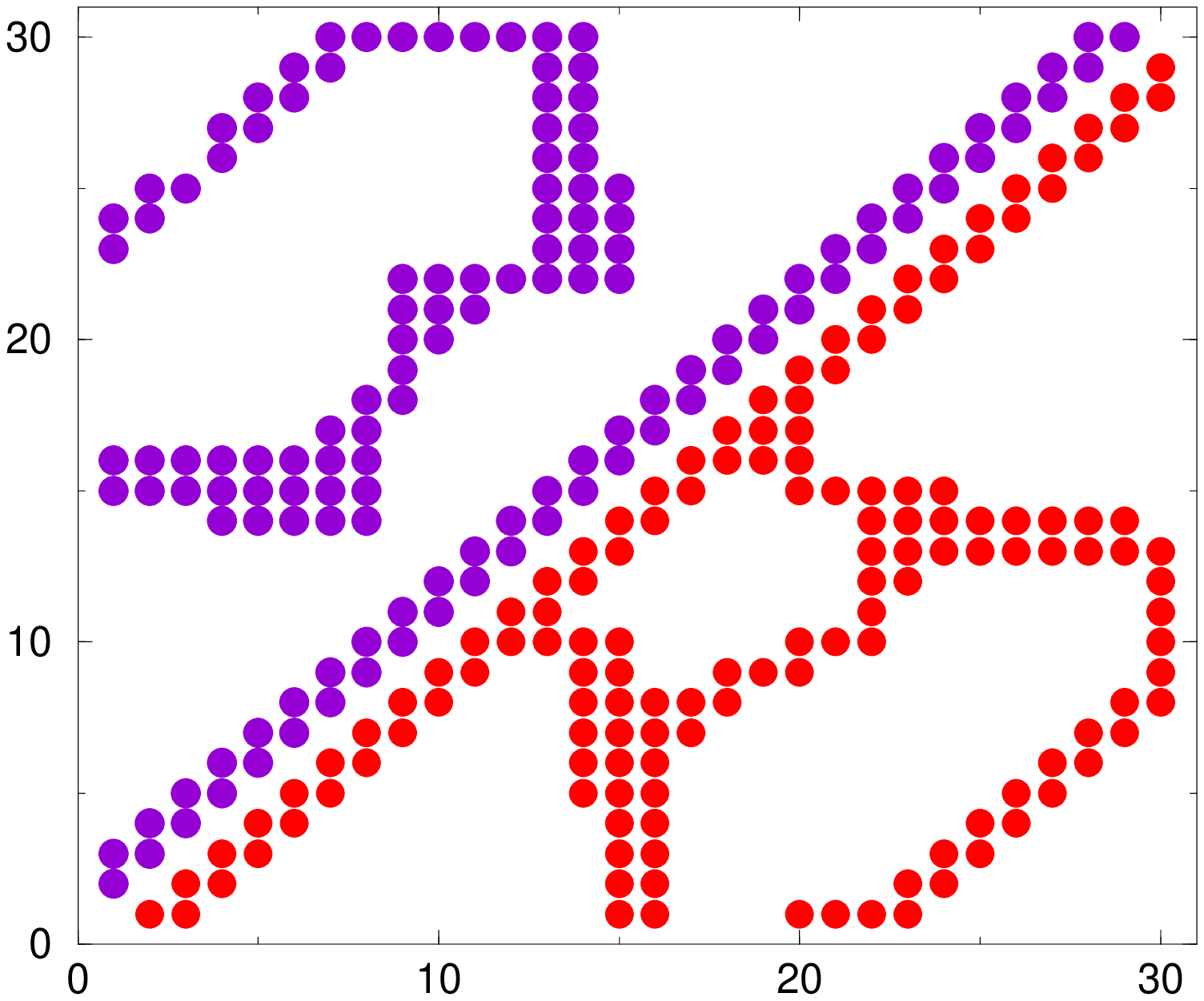,height=7.0cm,angle=0}}
\caption{
}
\label{fig:mseq5_pred}
\end{figure}

\begin{figure}
\centerline{\psfig{figure=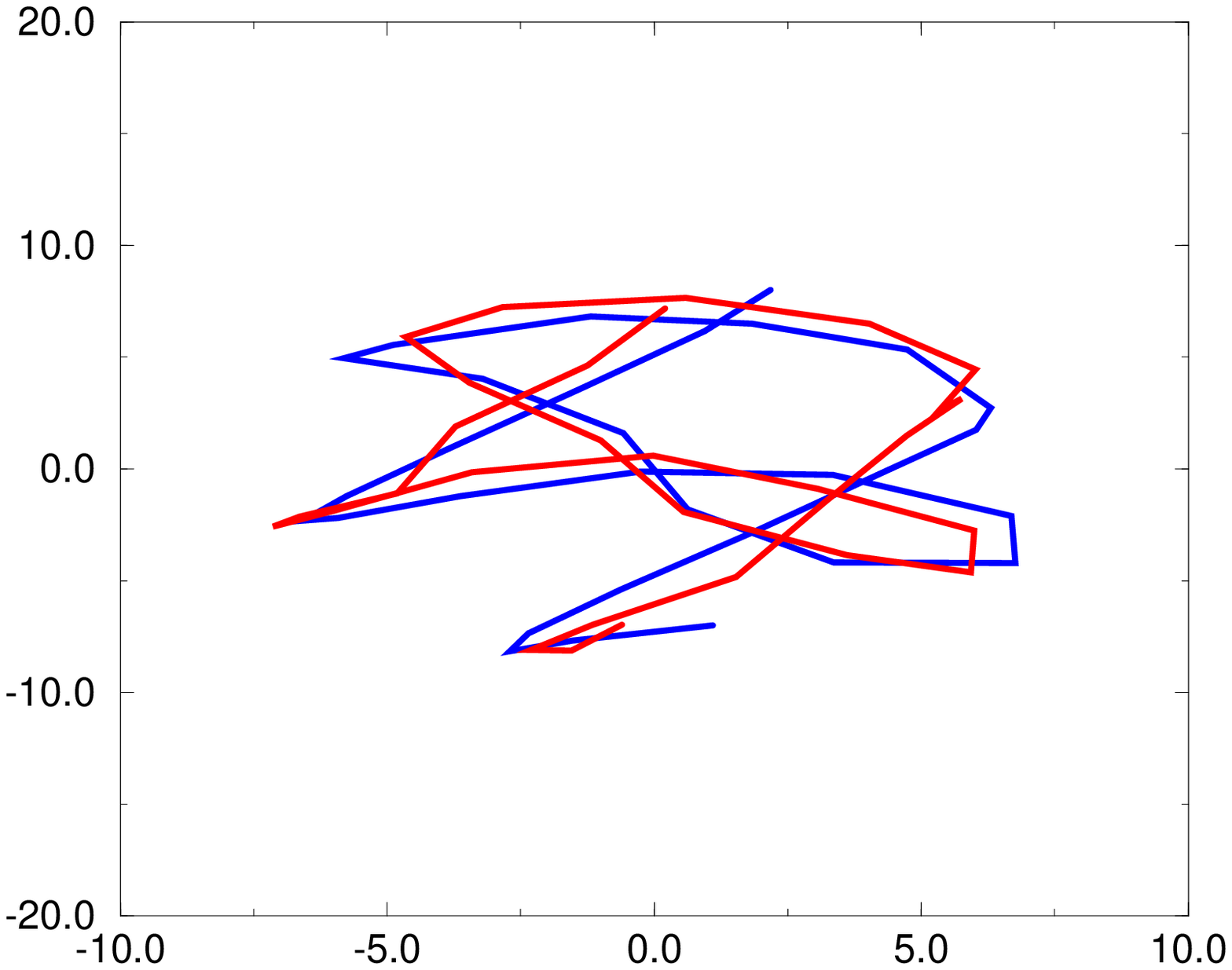,height=7.0cm,angle=0}}
\caption{ 
}
\label{fig:mseq5.bbc}
\end{figure}

\begin{figure}
\centerline{\psfig{figure=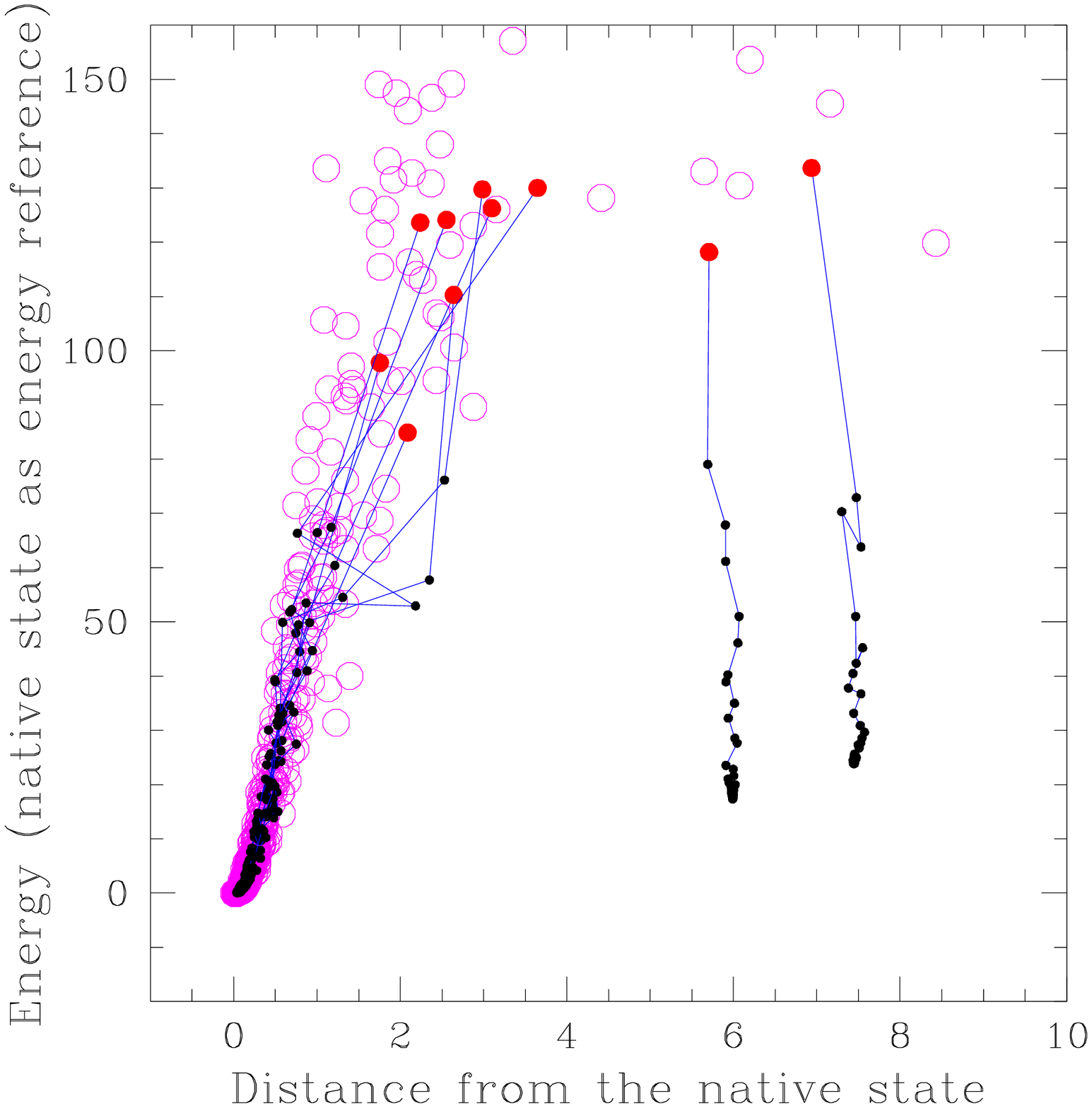,height=7.0cm,angle=0}}
\caption{
}
\label{fig:md_seq5}
\end{figure}

\end{document}